\RequirePackage{fix-cm}

\documentclass[12pt,journal,compsoc,onecolumn]{IEEEtran}   

\usepackage{setspace}
\usepackage{graphicx}
\usepackage{balance}
\usepackage{times}
\usepackage{color}
\usepackage{ifpdf}
\usepackage{subfigure}
\usepackage{epsfig}
\usepackage{latexsym}
\usepackage{amsfonts}
\usepackage{amssymb}
\usepackage{comment}
\usepackage{xspace}
\usepackage{mathrsfs}
\usepackage{amssymb}
\usepackage{color}
\usepackage{algorithmic}
\usepackage[ruled]{algorithm}
\usepackage{amssymb}
\usepackage{array}
\usepackage{url,epsfig,array}
\usepackage{leftidx}
\usepackage{amsmath}
\usepackage{setspace}
\usepackage{float} 
\usepackage{multirow}

\newtheorem{theorem}{Theorem}
\newtheorem{lemma}[theorem]{Lemma}

\def\ie{\textit{i.e.}\xspace}

\begin{document}
\doublespacing

\title{QUICK: QoS-guaranteed Efficient Cloudlet Placement in Wireless Metropolitan Area Networks\thanks{This work has been accepted by The Journal of Supercomputing on May 7, 2018. This manuscript is not the accepted version. Contents may vary from the final version. This manuscript is used for the purpose of quick disemination of research findings. Copyright are held by authors and corresponding copyright holders.}
}

	\author{\IEEEauthorblockN{Long Chen\IEEEauthorrefmark{1},
		Jigang Wu\IEEEauthorrefmark{2},  Gangqiang Zhou\IEEEauthorrefmark{3}, Lonjie Ma\IEEEauthorrefmark{4} }    \\
	\IEEEauthorblockA{\IEEEauthorrefmark{1}School of Computer Science and Technology\\
		Guangdong University of Technology,	Guangzhou 510006, Guangdong Province, China\\
			Email:\IEEEauthorrefmark{1}lonchen@mail.ustc.edu.cn, \IEEEauthorrefmark{2}asjgwucn@outlook.com
	}
}

\maketitle

\section{abstract}
This article defines the QoS-guaranteed efficient cloudlet deploy problem in wireless metropolitan area network, which aims to minimize the average access delay of mobile users \textit{i.e.} the average delay when service requests are successfully sent and being served by cloudlets. Meanwhile, we try to optimize total deploy cost represented by the total number of deployed cloudlets. For the first target, both un-designated capacity and constrained capacity cases are studied, and we have designed efficient heuristic and clustering algorithms respectively. We show our algorithms are more efficient than the existing algorithm. For the second target, we formulate an integer linear programming to minimize the number of used cloudlets with given average access delay requirment. A clustering algorithm is devised to guarantee the scalability. For a special case of the deploy cost optimization problem where all cloudlets' computing capabilities have been given, i.e., designated capacity, a minimal cloudlets efficient heuristic algorithm is further proposed. We finally evaluate the performance of proposed algorithms through extensive experimental simulations. Simulation results demonstrate the proposed algorithms are more than $46\%$ efficient than existing algorithms on the average cloudlet access delay. Compared with existing algorithms, our proposed clustering and heuristic algorithms can reduce the number of deployed cloudlets by about $50\%$ averagely.

\begin{IEEEkeywords}
Cloudlet, Access delay, Cloud computing, Heuristic, Clustering
\end{IEEEkeywords}


\section{Introduction}\label{sec:intro}
With the emerging of cloud computing and wireless communication technologies, mobile devices have become proliferation in people's daily life. Due to the small size nature of mobile devices, their computation resources are limited \cite{pang2015survey}. To improve user experience and enhance battery life, mobile applications can be offloaded to a remote cloud through wireless links \cite{zhang2012offload}. However, the cloud is usually far away from users, which results in a high end-to-end latency \cite{verbelen2012cloudlets}. To meet the delay requirment of mobile applications, cloudlets are proposed to extend the traditional mobile device-cloud architecture \cite{satyanarayanan2009case}. The cloudlet is a trusted, computation rich device that is deployed at the access point (AP) or base station \cite{pang2015survey}, available to be used by nearby mobile devices. For the mobile device, access proximity to the cloudlet is very important to reduce end-to-end delay. So the best location for the cloudlet is one wireless hop away from the mobile device. 

Although there have been extensive studies on resource allocation in cloudlet based networks, there are only a few works focusing on cloudlet deployment. Meanwhile, how to guarantee the quality of service (QoS) of cloudlet based wireless metropolitan area network (WMAN) is somehow overlooked. In hot spot areas \cite{fazio2017prediction}, cloudlet servers can always be accessed by a large number of mobile users whereas it will not be the case in sparsely populated areas and the cloudlets' resources may be wasted. If we allocate the cloudlet resources based on the number of user requests, the cloudlet utilization efficiency could be improved. Due to the limited budget of network operator, how to reduce deployment cost by reducing the number of cloudlets and how to choose proper locations for cloudlets are critical for the service provider \cite{xu2016efficient}. Although mobile users are dynamic and user requests changes frequently, we can determine where to deploy the cloudlet servers based on the data of historic statistics. During the cloudlet placement procedure, there are two main challenges. 1) Which cloudlet should be placed at which AP must be decided so that user requests can be properly handled within short delays. 2) How many cloudlets should be placed to satisfy bounded user delay requirements must be determined.

In this paper, we focus on the \underline{Q}oS-g\underline{u}aranteed eff\underline{i}cient \underline{c}loudlet placement problem in wireless metropolitan area networ\underline{k} (Quick). Classified by different research targets, the Quick problem then can be divided into two sub-problems, \ie, the \underline{q}uality \underline{o}f \underline{e}xperience oriented \underline{c}loudlet \underline{p}lacement (QOECP) problem and the \underline{d}elay \underline{b}ounded \underline{o}ptimal \underline{c}loudlet \underline{p}lacement and user association (DBOCP) problem.

The main contributions of this paper are summarized as follows:
\begin{itemize}
\item We study the QOECP problem in two cases \ie, un-designated capacity and constrained capacity. Then we devise two algorithms named MDC and MDE accordingly to efficiently deploy cloudlets aiming at improving the running performance of existing heuristic algorithm.
\item We then study the DBOCP problem in WMANs by showing the problem is NP-hard, and formulate it as an Integer Linear Programming (ILP).
\item Due to the poor scalability of the ILP, we devise a Minimal K Clustering algorithm (MKC) \cite{barioni2008accelerating} for the DBOCP problem without considering the capacity limit.
\item For a special case of the DBOCP problem where all cloudlets' computation capabilities have been given, we devise a minimal K heuristic algorithm (MKH) to minimize the number of cloudlets.
\item Simulation results demonstrate the proposed algorithms are more than $46\%$ efficient than existing algorithms on the average cloudlet access delay. Moreover, compared with existing algorithms, our proposed clustering and heuristic algorithms can reduce the number of deployed cloudlets by about $50\%$ averagely.
\end{itemize}

The rest of the paper is organized as follows. Section \ref{sec:related} reviews related work. Section \ref{sec:problem} introduces the system model and problem definitions. Section \ref{sec:Qoec} presents algorithms for the QOECP problem. Section \ref{sec:Dota} provides two algorithms to solve the DBOCP problem. Section \ref{sec:experiment} evaluates the performance of proposed algorithms by experimental simulations, and Section \ref{sec:conclude} concludes this paper.

\section{Related Work}\label{sec:related}
Cloud based task offloading has been extensively studied in the literature, e.g. \cite{Ren2014Dynamic} \cite{Gu2014Optimal} \cite{kosta2012thinkair}. Due to the long average access delay between users and remote clouds, cloudlets have been proposed as an alternative offloading technique \cite{Xu2015Capacitated}. In \cite{jia2015optimal}, authors investigate the cloudlet placement problem to minimize the response time of user requests. In \cite{cai2014cloudlet}, a game based ad hoc cloudlet network architecture is proposed, where server transmission rate can be significantly reduced. However, those solutions are suitable for merely a small number of user requests. Despite the increasing cloudlet research, the question on where to deploy cloudlets in a network has been overlooked \cite{xu2016efficient} \cite{Xu2015Capacitated}. In \cite{bourdena2015using}, a novel architecture creates a framework that enables ad hoc discovery of nearby devices in the network to share resources. But there are serious security and privacy concerns when offloading tasks to ad hoc devices. Therefore, similar to \cite{jia2015optimal} in this paper, cloudlets are adopted, deployed at the AP.

Cloudlets can be used as an infrastructure deployed in the existing WMAN network \cite{hoang2012optimal}. The QOECP problem is similar to the facility location problem \cite{Cormen2001Introduction}, however, they are different in the following aspects. For the facility location problem, it consists of a set of potential facility sites where a facility can be deployed, and a set of demand points that must be serviced. The goal is to choose a subset of facilities to operate, to minimize the sum of distances from each demand point to its nearest facility, plus the sum of operating costs of the facilities. In a metropolitan area, each user can offload their tasks to neighbouring APs and then the APs can offload tasks to cloudlet servers, which makes it difficult to effectively apply traditional facility location algorithms. The QOECP problem can also be reduced to the capacitated K-median problem and solved by the approximation algorithm \cite{Charikar1999A}. However, the approximation algorithm on a large-scale WMAN result in poor scalability when the network size is large.

There have been some studies focusing on placing cloudlets to reduce the average access delay between mobile users and the cloudlets \cite{Fan2017Cost} \cite{jia2015optimal} \cite{xu2016efficient}. In \cite{xu2016efficient}, authors propose a novel scheme to place a given number of cloudlets with different computing capacities to some strategic locations in WMAN with the objective to minimize the average access delay between mobile users and the cloudlets serving users. However, it is not efficient because of the repeated sorting process. In \cite{jia2015optimal}, authors study the joint cloudlet placement and user-cloudlet association problem in WMANs to reduce system response time considering the cost to deploy cloudlets in a given network. However, they all assume that cloudlets' capacity has already known a priori. Few study pays attention to the deploy cost minimization problem \textit{i.e.}, impact on the number of cloudlets with user access delay requirements. In \cite{Fan2017Cost}, the cloudlet placement for big data processing problem is addressed where the problem is modeled by two sub-problems,  \textit{i.e.} the cost minimization sub-problem and the access delay minimization sub-problem. However, they do not consider the QoS demands initiated by mobile users. 
\section{Problem Formulation}\label{sec:problem}
\subsection{System model}
\begin{table*}[htb]
	\caption{Symbols}\label{tb:sym}
	\centering
		\footnotesize
	\begin{tabular}{c|c}
		\hline
		Symbol & Definition\\
		\hline
		$G = (V\cup S, E)$ & a WMAN that consists of a set $V$ of APs and a set $S$ of potential locations for cloudlets, $S\subseteq V$\\
		$n=|V|, \epsilon=|E|$ & the numbers of APs in $V$ and links in $E$ \\
		$v_j$ & the $j$th AP in set $V$, $j\in [1,n]$ \\
		$R_j$ and $r_{m_j}$ & a set of user requests at AP $v_j$ and a user request in the set \\
		$\gamma_{m_j}$ & computing resource demand of user request $r_{m_j}$, $1\le m \le |R_j|$ \\
		$\omega(v_j)$ & an integer weight of AP $v_j$ that represents the number of user requests at $v_j$ \\
		$K$ and $C_i$ & number of cloudlets need to be placed and the $i$th cloudlet,  $(1\leq i\leq K)$ \\
		$c_1, c_2,...,c_K$ & the computing capacities of the $K$ cloudlets\\
		$d_{jl}$ & the service delay for the cloudlet at location $v_l$ serving the client at location $v_j$\\
		$x_{lm_j}$ & a binary indicator variable that show whether request $r_{m_j}\in R_j$ is assigned to cloudlet at $v_l$\\
		$z_{jl}$ & number of user requests at AP $v_j$ that are sent to cloudlet at location $v_l$ \\
		$p_{il}$ & a binary variable, where $p_{il}=1$ implies that cloudlet $C_i$ is placed at location $v_{l}$\\
		$D$ & given average cloudlet access delay\\
		$D_{avg}$ & average cloudlet access delay of all user requests\\
		$D_{ij}$ & sum of delay of the requests assigned to the cloudlet $C_i$ at $\upsilon_j$\\
		$R_{tot}$ & total user requests at APs\\
		$D_{tot}$ & sum of delay of the all requests assigned to the cloudlet\\
		$\gamma_{sum}$ & sum of computing resource of the all requests assigned to the cloudlet\\
		$L$ & the set of deployed locations of cloudlets  \\
		$d(e)$ & the communication delay of each edge $e\in E$ \\
		\hline
	\end{tabular}
\end{table*}
We consider a WMAN $G = (V \cup S, E)$ a set $V$ of APs, a set $S$ of potential cloudlets locations, and a set $E$ of links between two APs. Where $n=|V|$ and $\epsilon=|E|$. Let $R_j$ be the set of user requests at AP $v_j$. Denote $r_{m_j} \in R_j $ as user request with computing resource demand $\gamma_{m_j}$ and $w(v_j)=\sum_{m=1}^{|R_j|}r_{m_j}$ be the number of requests at $v_j$, where $m\in [1,|R_j|]$. Noting that APs are usually deployed at specific locations such as shopping malls, bus stations, schools, libraries, etc., so the number of user requests at each AP per unit time can be estimated by the population density in that area, or the historic AP access information through a linear regression technique. For detail notations see Table \ref{tb:sym}.
\begin{figure}[htb]
	\centering
	\setlength{\belowcaptionskip}{-1em}
	\includegraphics[width=3.0in]{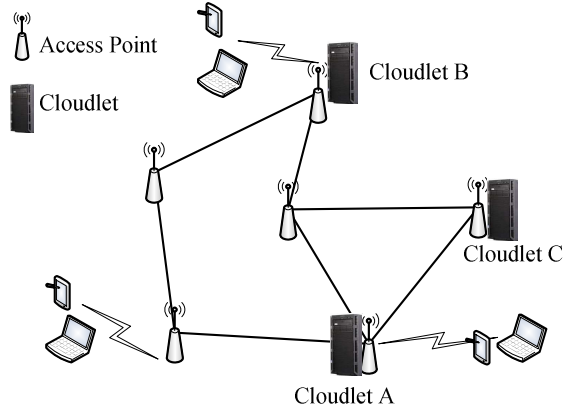}
	\caption{A WMAN $G = \langle V \cup S, E \rangle$ with $K = 3$ cloudlets.}\label{fig:network}
\end{figure}

As shown in Fig. \ref{fig:network}, we assume cloudlets are co-located with APs. There are $K$ cloudlets needed to be placed in $K$ different potential locations in $S$. For each cloudlet $C_i$, it has a limited computing resource to process user requests. Let $c_i$ be the computing capacity of $C_i$. Mobile users can offload their tasks to local cloudlets via the communication with APs. Therefore, if the current cloudlet can serve neighbouring user requests close to the AP it attaches to, user access delay can be reduced; otherwise, the user requests at that AP must be relayed to nearby cloudlets. We assume, all nodes stay staic during the allocation period \cite{Jin2016Auction} and the cloudlets can be initially deployed based on the statistical history data. To deal  the original resquest at $v_i$ can be redirected to other APs.

\subsection{QOECP problem and DBOCP problem Definitions}
For the QOECP problem, given an integer $1 \le K \le |S|$, place $K$ cloudlets in the WMAN co-located with some APs and assign users to the cloudlets to minimize the average cloudlet access delay between mobile users and cloudlets. For the un-designated capacity case, the set $\{C_1, C_2,...,C_K\}$ of $K$ cloudlets are with abundant computation resources available to be placed into $K$ different locations. For the constrained capacity case, capacities of $K$ cloudlets have already been given, \ie cloudlets with capacities $c_1\leq c_2 \leq,...,\leq c_k\leq \gamma_{sum}$.

For the DBOCP problem, given a tolerate delay $D\geq 0$, place $K$ cloudlets at some APs and assign user requests to the cloudlets, the objective is to minimize cloudlets number $K$. The un-designated capacity case and the constrained capacity case are the same as QOECP problem despite of the different solution targets.

\subsection{NP-Hardness}
The QOECP problem has been proved to be NP-hard in \cite{xu2016efficient}. We show DBOCP problem is NP-hard by reduction from the set covering problem, which is NP-hard.
\begin{lemma}
The QOECP problem in $G=\langle V\bigcup S, E\rangle$ is NP-hard.
\begin{proof}
In the set covering problem, given two sets $U$ and $O$, where $U$ is the universe set and $O \subseteq U$. Take a set $C \subseteq O$ and make the union of $C$ equal to $U$. For the set covering optimization problem, the input is a pair (U,O) with an integer $t\le|C|$. The task is to find a set covering that uses the fewest sets, i.e. to minimize $t$ \cite{Cormen2001Introduction}. The DBOCP problem is to place $K$ cloudlets to the poteintial location set $S$, such that the number of deployed cloudlets $K$ is minimized which is equivalent to determine a subset of cloudlets $C \subseteq S$, make the union in $C$ equal to $S$ and minimize $|C|$, which concludes the proof. 
\end{proof}
\end{lemma}

\subsection{Formulation of QOECP and DBOCP Problems}
The QOECP problem can be formulated as an ILP. We use a binary variable $p_{il}$ to indicate whether cloudlet $C_i$ will be placed to location $\upsilon_{l}\in S$ or not. Where $p_{il} = 1$ if cloudlet $C_i$ is placed at $\upsilon_{l}$ and $p_{il} = 0$ otherwise, for $\forall i\in [1,K]$ and $\forall l\in [1,|S|]$. Similarly, we use a binary variable $x_{im_j}$ to indicate whether a user request $r_{m_j}\in R_{j}$ will be assigned to a cloudlet located at $\upsilon_{i}$ or not. That is, $x_{lm_j}=1$ if $r_{m_j}$ is assigned to the cloudlet at location $\upsilon_{l}$ and 0 otherwise. Let $z_{ij}$ be the number of user requests at AP $\upsilon_{j}$ that are assigned to the cloudlet at location $\upsilon_{i}$. Clearly, $\sum_{r_{m_j}\in R_{j}}x_{lm_j}=z_{jl}$. Note that, in an AP, different user requests may be assigned to different cloudlets. Let $d_{ij}$ be the length of a shortest path between $\upsilon_{i}$ and $\upsilon_{j}$ in terms of accumulated delay, then we have \cite{xu2016efficient},

\textit{QOECP \-- Problem}
\begin{equation}
\begin{split}
\centering
&  obj: \min \frac{\sum_{j=1}^{|V|}\sum_{l=1}^{|S|}z_{jl}d_{jl}}{\sum_{i=1}^{|V|}\omega(\upsilon_{i})}
\\
 \quad \quad \quad s.t. & \sum_{l=1}^{|S|}p_{il}=1, \forall  1\leq i\leq K \quad \quad  \quad  \quad \ \ \ \  \quad \quad \quad \ (C1) \\   
 \quad \quad \quad &\sum_{l=1}^{|S|}p_{il}\leq 1, \forall  \upsilon_{l}\in S \quad \quad \quad  \quad  \quad \ \ \ \ \ \  \quad \quad \quad \ \ (C2) \\
 & \sum_{r_{m_j}\in R_{j}}x_{lm_j}=z_{jl}, \forall\upsilon_{j}\in V, \upsilon_{l}\in S  \ \  \quad \quad \quad \ (C3)\\
 &\sum_{l=1}^{|S|}z_{jl}=\omega(\upsilon_{i}), \forall \upsilon_{i}\in V   \quad \quad \quad \quad \ \ \ \  \quad \quad \quad (C4)\\
 &  \frac{z_{jl}}{\omega(\upsilon_{l})}\leq \sum_{i=1}^{|K|}p_{il}, \forall\upsilon_{i}\in V, \forall \upsilon_{l}\in S  \quad  \  \quad \quad \quad \ (C5)\\
 & p_{il}\in \{0,1\}, x_{lm_j}\in \{0,1\}, \quad \quad \quad \quad \  \quad \quad \quad  (C6) \\
 &\sum_{j=1}^{n} \sum_{r_{m_j}\in R_j} \gamma_{m_j} \cdot x_{l_{m_j}} \le \sum_{i=1}^{K}p_{il}c_i, \forall v_l \in S . \ \ \  (C7)
\end{split}
\end{equation}
Where constraint $C1$ ensures that each cloudlet is deployed at only one location in $S$, and $C2$ ensures that at most one cloudlet is deployed at one location. Constraints $C3$ and $C4$ ensure that all user requests
from each AP $v_j$ will be assigned to cloudlets, while $C5$ ensures that whenever some user
requests at AP $v_i$ are assigned to location $v_l$, then one of the $K$ cloudlets must be placed at location $v_l$. Constraints in $C6$ and $C7$ are binary and capacity constraints.

Similarly, the DBOCP problem can be formulated as follows:

\textit{DBOCP \-- Problem}
\begin{equation}\label{eq:minK}
\begin{split}
\quad \quad \quad \quad \quad \quad & obj: \min   K \\
  s.t. \quad \quad & C2\--C7 \quad \quad and  \\
 & \frac{\sum_{j=1}^{|V|}\sum_{l=1}^{|S|}z_{jl}d_{jl}}{\sum_{i=1}^{|V|}\omega(\upsilon_{i})} \leq D,  \quad \quad \quad \quad \quad \quad (C8)   
\end{split}
\end{equation}
where constraint $C8$ in (\ref{eq:minK}) ensures that the average cloud access delay of the user requests in all APs $V$ no more than the given average cloudlet access delay $D$. The objective is to minimize the number of deployed cloudlets $K$.

\section{Algorithms for QOECP Problem}\label{sec:Qoec}
\subsection{ Minimal Delay Clustering Algorithm for the Un-designated Capacity Case}
In this subsection, we propose a  Minimal Delay Clustering algorithm (MDC) for the case when cloudlets' computing capabilities are not given, \ie un-designated capacity\footnote{Following \cite{xu2016efficient}, capacity means the ability to handle user requests which is represented by the maximum number of user requests one cloudlet can receive. In the following paragraphs, we do not rank capacity according to CPU cycles or data storage sizes one cloudlet can offer, as well as the case when one cloudlet might have a better CPU, while the other has more memory.}. In MDC, it first computes all shortest paths between each pair of APs in $G$. And then, for each AP $v_i$, it sorts in increasing order of delays between $v_i$ and the other AP $v_j$, where $ j\ne i$ and $j\in V$.

Inspired by the clustering algorithm \cite{barioni2008accelerating}, MDC first randomly selects $K$ location as the initial cloudlets' position. Then, it assigns user requests at AP $v_j$ to the nearest cloudlet $C_i$ at $v_i$ by dividing all APs into $K$ clusters and the cloudlet is positioned at the cluster $cl_i$'s center. Let $D_{ij}$ be the accumulated delay of $C_i$'s requests assigned to the $v_{j}$, update cluster $cl_i$'s center to minimize $D_{ij}$ following the swap and selection phases mentioned in \cite{barioni2008accelerating}. After completion of all clusters' forming and all clusters $cl_i$'s center being not changed, we can obtain the minimum average delay $D_{avg}$ in $K$ cloudlets. Finally, we can calculate the cloudlet $C_i$'s capacity $c_i$ according to assigned user requests. The detailed algorithm is depicted in Algorithms \ref{alg:MDC} and \ref{alg:clustering}.

\begin{theorem}
Given a WMAN $G=\langle V\cup S,E\rangle$, a set of user requests $R_j$ at each $\upsilon_j$ with $\omega(\upsilon_j)=R_{m_j}$, there is a fast, scalable K-Medoids \cite{barioni2008accelerating}  based algorithm for the QOECP problem, which takes $O(n\epsilon+Kn^2)$ time.
\begin{proof}
Let $n=|V|+|S|$ and $\epsilon=|E|$ be the number of nodes and edges in $G$ accordingly. Finding all pairs of shortest paths in $G$ takes time $O(n^{2}\log n+n\epsilon)$, by applying Dijkstra¡¯s algorithm for single-source shortest paths of all source nodes in $V \cup S$, the time complexity is $O(n\log n+\epsilon\log n)$ \cite{Cormen2001Introduction}. For all APs, sorting APs with to-be-allocated user requests in increasing order of their access delays between the AP and the other APs takes $O(n^2\log n)$ time. Dividing all APs into $K$ clusters takes $O(n)$ and assigning all the user requests in the cluster center takes $O(n^2)$, the changing and adjusting $K$ cluster centers takes $O(Kn^2)$. Therefore, total time complexity of Algorithm \ref{alg:MDC} is $O(n^2 \log n+n\epsilon+Kn^2)=O(n\epsilon+Kn^2)$.
\end{proof}
\end{theorem}

\begin{algorithm}[h]
\footnotesize	\begin{algorithmic}[1]
		\caption{ Minimal Delay Clustering Algorithm (MDC)}\label{alg:MDC}
\REQUIRE $G\langle V \bigcup S, E\rangle$, $R_j$, $\gamma_{m_j}$, $\omega(\upsilon_j)=R_{m_j}$, $d(e)$, $K$, $S(\subseteq V)$, $R_{tot}$, $D_{tot}$.
\ENSURE $L$, $c_1,c_2,\dots,c_k$,  $D_{avg}$.
\vspace{1ex}
\STATE $L\leftarrow \emptyset$;
\STATE Compute all pairs of shortest paths for each pair of APs in $G$;
\FOR{each potential location $\upsilon_l\in S$}
\STATE Sort the APs with to-be-allocated user requests in increasing order of access delays $d_{ij}$;
\ENDFOR
\STATE $D_{avg} \leftarrow 0$
\STATE $L$, $c_1,c_2,\dots,c_k$, $D_{avg}$ $\leftarrow$ $\textbf{Clustering}(G\langle V \bigcup S, E\rangle$, $R_j$, $\gamma_{m_j}$, $d(e)$, $S(\subseteq V)$, $D_{avg}$, $L$)
\RETURN $L$, $c_1,c_2,\dots,c_k$, $D_{avg}$.
\end{algorithmic}	
\end{algorithm}

\begin{algorithm}[h]
\footnotesize	\begin{algorithmic}[1]
		\caption{Clustering}\label{alg:clustering}
\REQUIRE $G\langle V \bigcup S, E\rangle$, $R_j$, $\gamma_{m_j}$, $d(e)$, $S(\subseteq V)$.
\ENSURE $L$, $c_1,c_2,\dots,c_k$,  $D_{avg}$.
\vspace{1ex}
\STATE $change\leftarrow true$
\STATE initialize $D_{tot}\leftarrow \emptyset$
\STATE Randomly select $K$ position as the initial cluster centers' (cloudlets') position.
\WHILE{$change$}
\STATE $D_{tot}\leftarrow 0$
\STATE Assign user requests at APs to the nearest cloudlet $C_i$, divide APs into $K$ clusters;
\FOR{each clusters $cl_i$}
\FOR{APs within cluster $cl_i$}
\STATE Assign all the user requests at the APs of each cluster $cl_i$, Let $D_{ij}$ be the sum of delays of the the clusters $cl_i$ requests assigned to the nearest cloudlet $C_{j}$, $j\ne i$;
\STATE Calculate each cloudlets' capacity at cluster center
\ENDFOR
\STATE Update the clusters $cl_i$'s center to obtain the minimum $D_{ij}$ for a AP $v_{j}, j\in V$ following the update and swap procedures in \cite{barioni2008accelerating},
\STATE  $D_{tot}\leftarrow D_{tot}+D_{ij}$
\IF{$K$ clusters center not change}
\STATE $change\leftarrow false$
\ENDIF
\ENDFOR
\FOR{each $cl_{j}$}
\STATE $L\leftarrow L\cup \{cl_{j}\}$, with $1\le j\le K$ where $v_j$ is the center of cluster $cl_j$
\ENDFOR
\STATE $D_{avg}\leftarrow D_{tot}/R_{tot}$
\ENDWHILE
\RETURN $L$, $c_1,c_2,\dots,c_k$, $D_{avg}$.
\end{algorithmic}	
\end{algorithm}

\subsection{Minimal Delay Efficient Heuristic Algorithm (MDE)}
We present a Minimal Delay Efficient heuristic algorithm (MDE) for the case when cloudlets' computing capabilities have been given. First, MDE sorts the $K$ cloudlets in decreasing order of their capacities, i.e., $c_1\geq c_2\geq \dots\geq c_K$. Then it computes all shortest paths between each pair of APs in $G$, i,e., between $v_i$ and $v_j$, where $i\in [1,n]$ and $j\neq i$. Different from \cite{pang2015survey}, to reduce repeated sorting process, after all pairs of shortest paths for each AP pairs have been computed, for each AP $v_i$ we first sort in increasing order of delays between the AP and others APs $v_j$, $j\ne i$. For example, in a network with 6 APs and 3 cloudlets needed to be deployed, when using Heuristic\footnote{The heuristic algorithm proposed by \cite{xu2016efficient}, where each cloudlet should be placed to a location that can cover as many user requests as possible.}, it sorts all APs with to-be-allocated user requests in increasing order of their access delays between AP and cloudlet  location for $6*3=18$ times, but by MDE, the sorting process of APs only takes 6 times. The detail of MDE is shown in Algorithm \ref{alg:MDEH} and Algorithm \ref{alg:cloudletPlace}. 

\begin{algorithm}[h]
\footnotesize	\begin{algorithmic}[1]
		\caption{Minimal Delay Efficient Heuristic Algorithm}\label{alg:MDEH}
\REQUIRE $G\langle V \bigcup S, E\rangle$, $R_j$, $\gamma_{m_j}$, $\omega(\upsilon_j)=|R_j|$, $d(e)$, $K$, $S(\subseteq V)$, $c_1,c_2,\dots,c_k$.
\ENSURE $L$,  $D_{avg}$.
\vspace{1ex}
\STATE $L\leftarrow \emptyset$;
\STATE Compute all pairs of shortest paths for each pair of APs in $G$;
\FOR{each potential location $\upsilon_l\in S$}
\STATE Sort APs with to-be-allocated user requests in increasing order of $d_{ij}$;
\ENDFOR
\STATE $D_{avg} \leftarrow 0$
\STATE $L\leftarrow$ $\textbf{cloudletPlace}$ ($G\langle V \bigcup S, E\rangle$, $R_j$, $\gamma_{m_j}$, $d(e)$, $S(\subseteq V)$,  $c_1,c_2,\dots,c_k$, $L$);
\RETURN $L$, $D_{avg}$.
\end{algorithmic}	
\end{algorithm}

\begin{algorithm}[!h]
\footnotesize	\begin{algorithmic}[1]
		\caption{cloudletPlace}\label{alg:cloudletPlace}
\REQUIRE $G\langle V \bigcup S, E\rangle$, $R_j$, $\gamma_{m_j}$, $d(e)$, $S(\subseteq V)$, $c_1,c_2,\dots,c_k$.
\ENSURE $L$, $D_{avg}$.
\vspace{1ex}
\STATE initialize $D_{tot}\leftarrow \emptyset$, $L\leftarrow \emptyset$;
\STATE Sort the $K$ cloudlets by their capacities in decreasing order;
\FOR{$i\leftarrow 1$ to $K$}
\STATE /*Place cloudlet $C_i$ with capacity $c_i$ to an unoccupied location*/
\STATE $U_i\leftarrow S\setminus L$; /* the set of potential locations */
\FOR{each potential location $\upsilon_j\in U_i$}
\STATE Find the first $r$ APs in the sorted AP sequence such that the sum of the resource demands of user requests in those APs are no less than $c_i$.
\STATE Assign user requests from the first $r-1$ APs to cloudlet $C_i$ at location $\upsilon_j$;
\STATE Sort requests at the $r$th AP in increasing order of their computing resource demands;
\STATE Allocate a subset of user requests at the $r$-th AP to $C_i$ until $c_i$ is met;
\STATE Let $D_{ij}$ be the sum of delays of the requests assigned to the cloudlet $C_i$ at location $\upsilon_j$;
\ENDFOR
\STATE Place cloudlet $C_i$ at the location $\upsilon_i$ with the minimum sum of delays, i.e.,$\upsilon_i=\arg\min_{\upsilon_l\in U_i}$\{$D_{ij}$\};
\STATE $D_{tot}\leftarrow$ $D_{tot}+D_{ij}$
\STATE $L\leftarrow L\cup \{\upsilon_i\}$;
\ENDFOR
\STATE $D_{avg}\leftarrow D_{tot}/R_{tot}$
\RETURN $L$,  $D_{avg}$.
\end{algorithmic}	
\end{algorithm}
\begin{theorem}
 Given a WMAN $G = \langle V \cup S, E\rangle$, $R_j$ at each $\upsilon_j$ with $\omega(\upsilon_j)=R_{m_j}$, and $K$ cloudlets $C_1,C_2,\dots,C_K$ with capacities $c_1,c_2,\dots,c_K$, respectively, there is a fast, scalable algorithm for the computation capacity constrained QOECP problem, with time complexity of $O(Kn^2 log n + n\epsilon)$, assuming that $\omega(\upsilon)\leq \min_{l\leq i\leq K}\{c_i\}$ for any $\upsilon \in V$ and $1\leq K\leq |S|\leq |V|$, where $n = |V|$, $\epsilon=|E|$ and $r$ is the index of the first AP that cannot be completely assigned to the current cloudlet.
 \begin{proof}
The solution delivered by Algorithm \ref{alg:MDEH} is a feasible solution since all user requests at each AP $\upsilon_j\in V$ are assigned to cloudlets and the number of user requests assigned to each cloudlet $C_i$ is no more than its capacity $c_i$.

Let $n=|V|+|S|$ and $\epsilon=|E|$. Finding all pairs of shortest paths in $G$ takes time $O(n^{2}\log n+n\epsilon)$, by applying Dijkstra¡¯s algorithm for single-source shortest paths of all source nodes in $V \cup S$, where the time complexity of Dijkstra¡¯s algorithm is $O(n\log n+\epsilon\log n)$\cite{Cormen2001Introduction}. Lines 3 to 5 in Algorithm \ref{alg:MDEH} takes $O(n^2\log n)$ time. Algorithm 3 proceeds iteratively and one of the $K$ cloudlets will be placed within each iteration. When placing cloudlet $C_i$ within iteration $i$, identifying a location $\upsilon_i\in S\setminus\{\upsilon_1,\upsilon_2, \dots, \upsilon_{i-1}\}$ for cloudlet $C_i$ placement takes $O(K\cdot |S|\cdot |r|\log |r|)=O(Knr\log r)$ time as $|S|\leq |V|=n$, $r\leq n$, $r$ is the index of the first AP that cannot be completely assigned to the current cloudlet. The total time complexity is $O(n^2 \log n+n\epsilon+n^2 \log n+Knr\log r)$, thus is no more than $O(Kn^2 log n + n\epsilon)$ and is faster than the Heuristic algorithm mentioned in \cite{xu2016efficient}.
\end{proof}
\end{theorem}

\section{Algorithms for DBOCP Problem}\label{sec:Dota}
\subsection{Minimal K Clustering Algorithm}
Similarly as QOECP problem, firstly we develop a Minimum number of $K$ Clustering algorithm (MKC) for the un-designated capacity case. MKC firstly computes all pairs of shortest paths for each pair of APs in $G$. And then, for each AP $v_i$ it sorts in increasing order of delays between the AP and other APs $v_j$. Recall that, in a WMAN $G$ the more number of cloudlets, the less average user requests delay. So MKC can gradually increase the number of cloudlets from $K=1$ to $n$, and find locations for cloudlets, making sure that all user requests assign to the cloudlets can meet their delay requirements.

MKC algorithm finds the location for cloudlet placement as follows. For an obtained $K$, we should have the minimum average cloudlet access delay. So we can use Algorithm \ref{alg:MDC}'s  process to place cloudlets. After completion of all clusters, we can obtain the minimum $D_{avg}$ of $K$ cloudlets, when $D_{avg}\leq D$, the cloudlets number $K$ is obtained. The detail MKC algorithm is depicted in Algorithm \ref{alg:MKC}.
\begin{theorem}
Given a WMAN $G=\langle V\cup S,E\rangle$, a set $R_j$ of user requests at each AP $\upsilon_j\in V$ with $\omega(\upsilon_j)=R_{m_j}$, there is a fast, scalable K-Medoids algorithm for the Dota problem, which takes $O(n\epsilon+\max\{K^2n^2, n^2\log n \})$ time, assuming that $\omega(\upsilon)\leq \min_{l\leq i\leq K}\{c_i\}$ for any $\upsilon \in V$ and $1\leq K\leq |S|\leq |V|$, where $n = |V|$, $\epsilon = |E|$ and $r$ is the index of the first AP that cannot be completely assigned to the current cloudlet.
\begin{proof}
The solution obtained by Algorithm \ref{alg:MKC} is a feasible solution since all user requests at each AP $\upsilon_j\in V$ are assigned to cloudlets, the number of cloudlets $K$ is minimized in a given average cloudlet access delay $D$. In each number of $K$ the algorithm finds the location for cloudlets making sure that all user requests assigned to cloudlets are within $D_{avg}$ bound and the cloudlet $C_i$'s capacity $c_i$ can be calculated. In the following, we analyze the time complexity of Algorithm \ref{alg:MKC}.

Let $n=|V|+|S|$ and $\epsilon=|E|$. Finding all pairs of shortest paths in $G$ takes $O(n^{2}\log n+n\epsilon)$, by applying Dijkstra's algorithm for single-source shortest paths for all source nodes in $V \cup S$, the time complexity of is $O(n\log n+\epsilon\log n)$\cite{Cormen2001Introduction}. Sorting APs with to-be-allocated user requests in increasing order of their access delays between the AP and the other APs takes $O(n^2\log n)$ time. Dividing all APs into $K$ clusters takes $O(n)$ and in the $K$ clusters assigning all the user requests to the cluster center takes $O(n^2)$, the changing and adjusting $K$ cluster centers takes $O(Kn^2)$ and there are at most $K$ rounds of repeated running of Algorithm \ref{alg:clustering}. The total time complexity of Algorithm \ref{alg:MKC} is thus $O(n^2 \log n+n\epsilon+K^2n^2)=O(n\epsilon+\max\{K^2n^2, n^2\log n \})$.
\end{proof}
\end{theorem}

\begin{algorithm}[h]
\footnotesize	\begin{algorithmic}[1]
		\caption{Minimal K clustering Algorithm (MKC)}\label{alg:MKC}
\REQUIRE $G\langle V \bigcup S, E\rangle$, $R_j$, $\gamma_{m_j}$, $\omega(\upsilon_j)=|R_j|$, $d(e)$, $K$, $S(\subseteq V)$, $D$, $R_{tot}$, $D_{tot}$.
\ENSURE The  minimize cloudlets numbers $K'$, $L$, $c_1,c_2,\dots,c_k$.
\vspace{1ex}
\STATE $L\leftarrow \emptyset$;
\STATE $K'\leftarrow \emptyset$;
\STATE Compute all pairs of shortest paths for each pair of APs in $G$;
\FOR{each potential location $\upsilon_l\in U$}
\STATE Sort APs with to-be-allocated user requests in increasing order of access delays $d_{ij}$;
\ENDFOR
\STATE $D_{avg} \leftarrow D+1$
\WHILE{$D_{avg}\geq D$ and $K'\le$ the total number of cloudlets $K$ }
\STATE $K'\leftarrow K'+1$
\STATE $L\leftarrow \emptyset$;
\STATE $L$, $c_1,c_2,\dots,c_k$ $\leftarrow$ $\textbf{Clustering}(G\langle V \bigcup S, E\rangle$, $R_j$, $\gamma_m$, $d(e)$, $S(\subseteq V)$, $D_{avg}$, $L$)
\ENDWHILE
\RETURN $L$, $K'$, $c_1,c_2,\dots,c_k$.
\end{algorithmic}	
\end{algorithm}

\subsection{Minimal K Heuristic Algorithm}
A Minimal $K$ Heuristic Algorithm (MKH) for constrained capacity case is proposed. First, MKH computes all pairs of shortest paths for each pair of APs in $G$. After all pairs of shortest paths for each pair of APs in $G$ have been computed it sorts each AP to other APs in increasing order of their access delays. Then MKH gradually increases the number of cloudlets from $K=1$ to $n$. For $K$, it first sorts the $K$ cloudlets in decreasing order of their capacities, i.e., $c_1\geq c_2\geq \dots\geq c_K$ and making sure that all user requests are assigned to cloudlets and the average cloudlet access delay $D_{avg}$ is bounded. To minimize $D_{avg}$ in each number of $K$, iteration process is utilized to find the location for cloudlets. The detailed algorithm is shown in Algorithm \ref{alg:MKH}.
\begin{algorithm}[h]
\footnotesize	\begin{algorithmic}[1]
		\caption{Minimal K Heuristic Algorithm}\label{alg:MKH}
\REQUIRE $G\langle V \bigcup S, E\rangle$, $R_j$, $\gamma_{m_j}$, $\omega(\upsilon_j)=|R_j|$, $d(e)$, $K$, $S(\subseteq V)$£¬ $c_1,c_2,\dots,c_k$, $D$.
\ENSURE The  minimize cloudlets numbers $K'$, $L$.
\vspace{1ex}
\STATE $L\leftarrow \emptyset$;
\STATE Compute all pairs of shortest paths for each pair of APs in $G$;
\FOR{each potential location $\upsilon_l\in U$}
\STATE Sort the APs with to-be-allocated user requests in increasing order of their access delays $d_{ij}$ between AP $v_i$ and all others AP $\upsilon_j$;
\ENDFOR
\STATE $K\leftarrow 0$
\STATE $D_{avg} \leftarrow D+1$
\WHILE{$D_{avg}\geq D$ and $K'\le$ the total number of cloudlets $K$}
\STATE $K'\leftarrow K'+1$
\STATE $L\leftarrow \emptyset$;
\STATE $L\leftarrow$ $\textbf{cloudletPlace}$ ($G\langle V \bigcup S, E\rangle$, $R_j$, $\gamma_{m_j}$, $d(e)$, $S(\subseteq V)$,  $c_1,c_2,\dots,c_k$, $D_{avg}$, $L$);
\ENDWHILE
\RETURN $L$, $K'$,  $c_1,c_2,\dots,c_k$.
\end{algorithmic}	
\end{algorithm}
\begin{theorem}
Given a WMAN $G=\langle V\cup S,E\rangle$, a set $R_j$ of user requests at each AP $\upsilon_j\in V$ with $\omega(\upsilon_j)=R_{m_j}$, there is a fast, scalable algorithm for the identical capacities Dota problem, which takes $O(K^2n^2\log n+n\epsilon)$ time, assuming that $\omega(\upsilon)\leq \min_{l\leq i\leq K}\{c_i\}$ for any $\upsilon \in V$ and $1\leq K\leq |S|\leq |V|$.
\begin{proof}
The solution by Algorithm \ref{alg:MKH} is a feasible solution since all user requests at each AP  are assigned to cloudlets and the number of user requests assigned to each cloudlet $C_i$ is no more than $c_i$. $K$ is minimized iteratively with given delay $D$. The algorithm finds the location for cloudlets making sure that all user requests can be assigned to the cloudlet and $D_{avg}$ is minimal. Let $n=|V|+|S|$ and $\epsilon=|E|$. Finding all pairs of shortest paths takes $O(n^{2}\log n+n\epsilon)$, by applying Dijkstra's algorithm for single-source shortest paths of all source nodes in $V \cup S$, the time complexity is $O(n\log n+\epsilon\log n)$\cite{Cormen2001Introduction}. Sorting APs in increasing order of user access delays between the AP and neighbouring APs takes $O(n^2\log n)$. Algorithm \ref{alg:MKH} proceeds iteratively and one of the $K$ cloudlets will be placed within each iteration. When it places cloudlet $C_i$ within iteration $i$, identifying a location $\upsilon_i\in S\setminus\{\upsilon_1,\upsilon_2, \dots, \upsilon_{i-1}\}$ for cloudlet $C_i$ takes $O(K\cdot |S|\cdot |r|\log |r|)=O(Knr\log r)$ time as $|S|\leq |V|=n$, $r\leq n$ and there are at most $K$ rounds of repeated running of Algorithm \ref{alg:cloudletPlace}. The total time complexity of Algorithm \ref{alg:MKH} is $O(n^2 \log n+n\epsilon+K^2nr\log r)$, thus is no more than $O(K^2n^2\log n+n\epsilon)$.
\end{proof}
\end{theorem}

\section{Performance Evaluation}\label{sec:experiment}
\subsection{Experiment Settings}
The WMAN $G$ consists of $200$ to $1000$ APs and for each AP the probability that it is linked by another AP is set as 0.02 and the edge delay is randomly generated during $[5ms,50ms]$ following \cite{satyanarayanan2009case} and \cite{xu2016efficient}. The network topology is generated by a popular tool GT-ITM \cite{gtitm}. We assume $S = V$. The number of user requests $\omega(\upsilon)$ at each AP is randomly drawn from an interval $[50, 500]$ as did in \cite{IEEEstd8021990}. In the synthetic network computing resource demand $\gamma_{m_j}$ of each user request $r_{m_j}$ varies from $50$MHz to $200$ MHz \cite{jia2015optimal}. Let $\gamma_{sum}$ be the total amount of computing resource demands of all user requests, then $\gamma_{sum}=\sum^{n}_{j=1}\sum_{r_{m_j}\in R_j}\gamma_{m_j}$.The sum capacities of $K$ cloudlets is no less than $\gamma_{sum}$. Unless otherwise specified, these default parameters will be adopted in our simulation and the simulation values are obtained and averaged via $100$ times repeated running.

For comparing purposes, We evaluate the performance of proposed four algorithms of MDC, MDE, MKC and MKH with algorithms Random, Top-K \cite{xu2016efficient} and the optimal solution named as OPT obtained by solving the integer linear programming using LP-SOLVE \cite{lpsolve}. For the QOECP problem, the Random algorithm places the $K$ cloudlets to APs randomly, the Top-K places the $K$ cloudlets to the top-K APs according to the increasing number of requests at APs. For the DBOCP problem, the Random increases the number of cloudlets from $K = 1$ to $n$ and in each case, place the $K$ cloudlets to APs randomly. Different from the Random algorithm, the Top-K algorithm places the $K$ cloudlets to the top $K$ APs according the increasing sequence of user requests at APs.

\subsection{Performance Evaluation of Algorithms for QOECP Problem}
\begin{figure}[t]
 \centering
 \includegraphics[width=3.5in]{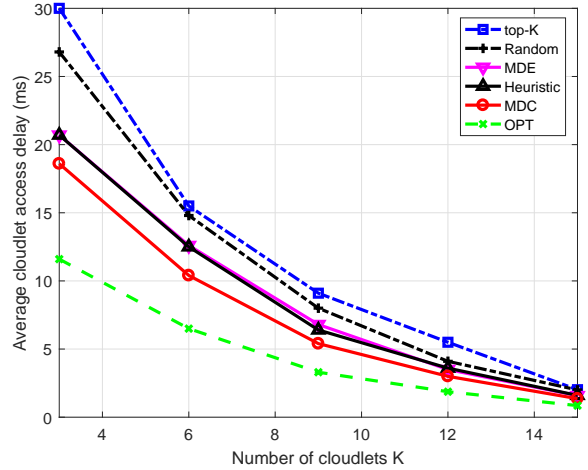}\caption{Average cloudlet access delay with number of cloudlets $K$ when the network size is small.}\label{fig:delayvsKsmall}
\end{figure}

\begin{figure}
	\centering
	\includegraphics[width=3.5in]{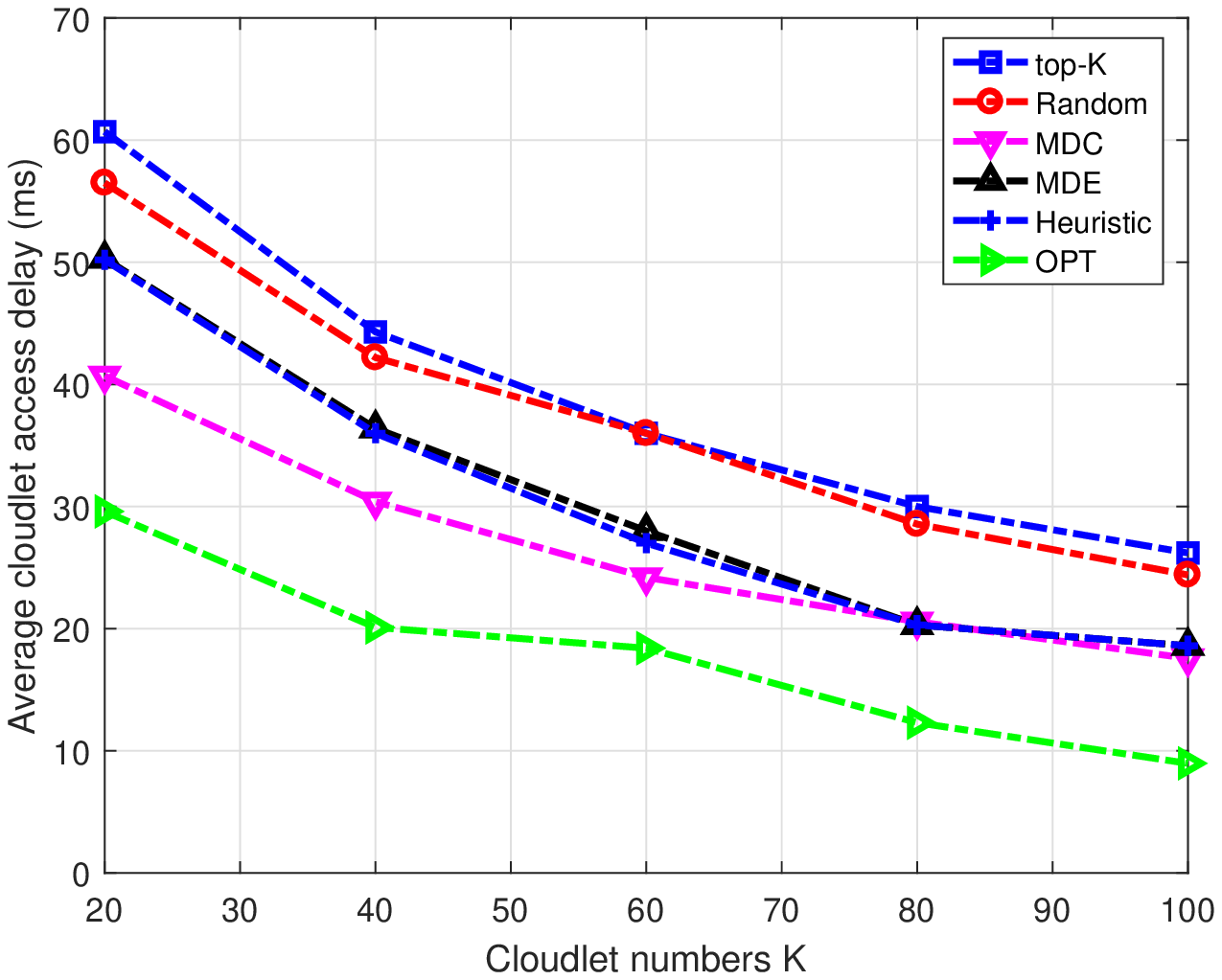}\caption{Average cloudlet access delay with number of cloudlets $K$ when the network size is large.}\label{fig:delayvsKlarge}
\end{figure}

\begin{figure}[t]
	\centering
	\includegraphics[width=3.5in]{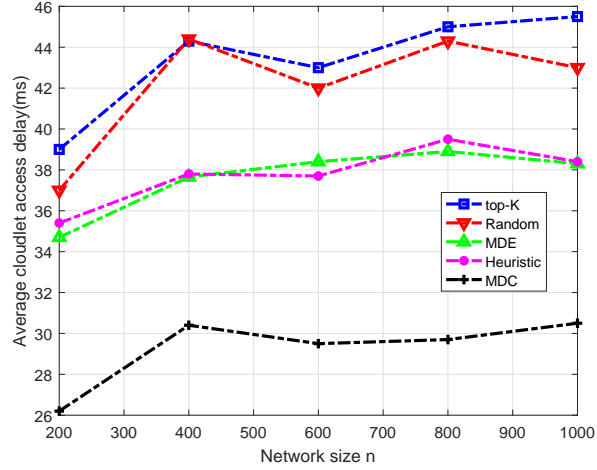}\caption{Average cloudlet access delay with constant number of cloudlets ($K=10\%\times n$) when network size is large.}\label{fig:avgdelayvsnconsk}
\end{figure}
\begin{figure}[t]
	\centering
	\includegraphics[width=3.5in]{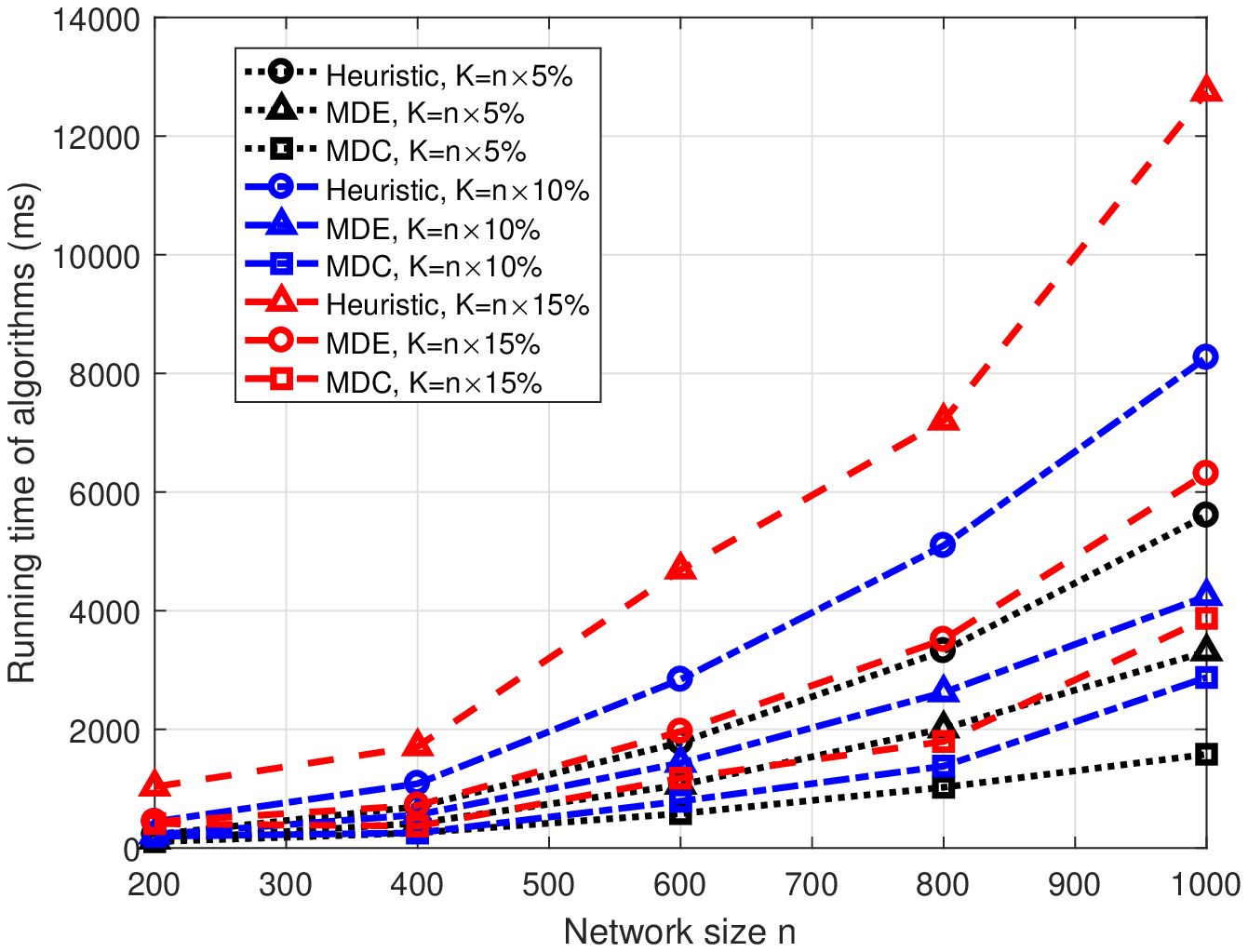}\caption{Running time of algorithms with number of APs}\label{fig:rtvsapnummdec}
\end{figure}
Firstly, we evaluate the performance of algorithms MDE, MDC with top-K, Random, Heuristic and OPT by setting the total number of APs as $n=18$. As shown in Fig. \ref{fig:delayvsKsmall}, for all algorithms, with the increasing number of $K$, the average cloudlet access delay decreases fast and the OPT algorithm achieves the best access performance while top-K performs the worst. On average, MDC outperforms MDE and Heuristic by about $15.28\%$ and MDC outperforms Random and Top-K by $46.77\%$ and $71.56\%$.  In addition, the figure also shows that the average cloudlet access delay for MDE is no more than 1.6 times of the optimal one. 

Then we observe the relationship between average cloudlet access delay and the number of cloudlet when the network size is large, as shown in Fig. \ref{fig:delayvsKlarge}, which demonstrates similar results as that of Fig. \ref{fig:delayvsKsmall}. The average access delay for all algorithms in the large size network is no more than 1.5 times of the cases in the small size network. 

In the third experiment, we fix the number of cloudlets as $K=10\% \times n$ and examine the average cloudlet access delay performance with the increasing number of APs larger than $200$. From Fig. \ref{fig:avgdelayvsnconsk}, we can see that MDC algorithm is  about $1.3$ times better than MDE and Heuristic algorithms. Moreover, MDC significantly outperforms Random and Top-K, and Random is only marginally better than algorithm Top-K. For the OPT algorithm, it fails to obtain the average cloudlet access delay due to high computation complexity. 

In the fourth experiment, we examine the running time of proposed algorithms with network size. Fig. \ref{fig:rtvsapnummdec} plots the curves of the running time delivered by Heuristic, MDE and MDC with changing network size and different number of cloudlets. For all algorithms, with network size grows, running time of algorithms will increase exponentially when $K$ is fixed. Moreover, when $n$ stays constant, with $K$ increases, the running time will also increase, which means algorithm running time is proportional to $K$ and $n$. On average, MDE is about 51.4\% better than the Heuristic algorithm because it reduces the repeated sorting process of APs. Among the poposed three algorithms, MDC shows the best performance and it is about $1.3$ times efficient than MDE.

\subsection{Performance Evaluation of Algorithms for DBOCP Problem}
\begin{figure}
\centering
\includegraphics[width=3.5in]{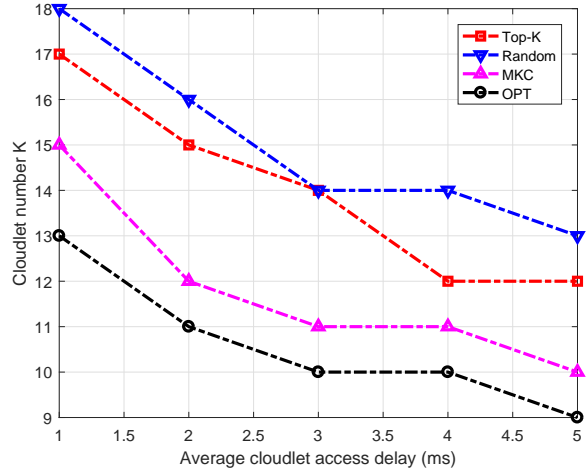}\caption{$K$ with average cloudlet access delay 
	$D$ with un-designated capacity. }\label{fig:KvsAvgDudc}
\end{figure}
\begin{figure}
	\centering
	\includegraphics[width=3.5in]{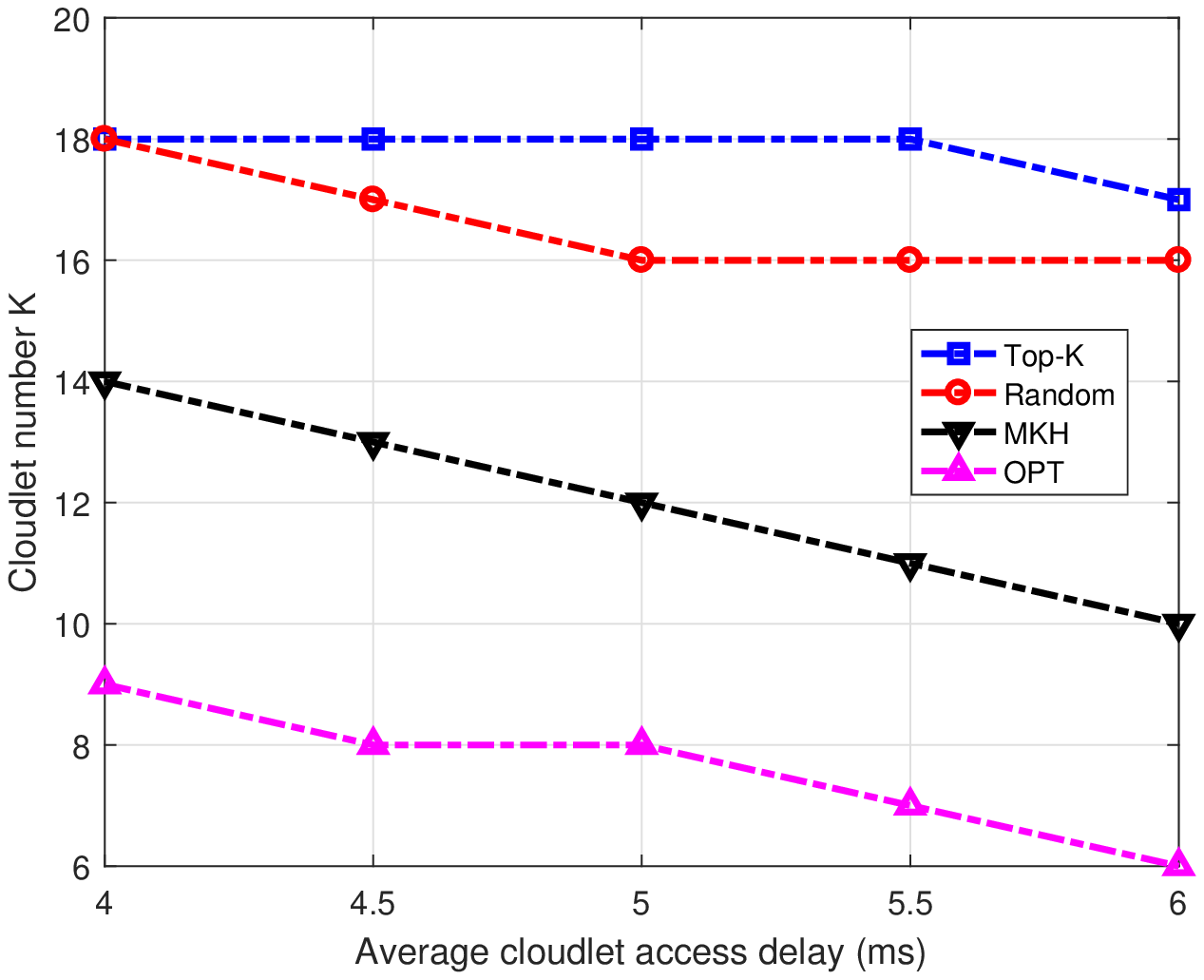}\caption{$K$ with average cloudlet access delay 
		$D$ with designated capacity. }\label{fig:KvsAvgDwdcc}
\end{figure}

\begin{table*}
	\small
	\caption{The solution $K$ and the algorithm runtime $T$ (ms), OPT with MKH}	\label{table:T2}
	\centering
	\begin{tabular}{l|ll|ll|ll|ll|ll|ll}
		\hline
		\multirow{3}{*}{n}
		&\multicolumn{4}{c|}{D=20}&\multicolumn{4}{c|}{D=25}&\multicolumn{4}{c}{D=30}\\
	\cline{2-5}\cline{6-9}\cline{10-13}
	&\multicolumn{2}{l|}{OPT}&\multicolumn{2}{l|}{MKH}&\multicolumn{2}{l|}{OPT} &\multicolumn{2}{l|}{MKH}&\multicolumn{2}{l|}{OPT}&\multicolumn{2}{l}{MKH}\\
	\cline{2-3}\cline{4-5}\cline{6-7}\cline{8-9}\cline{10-11}\cline{12-13}
		&$K$&$T$&$K$&$T$&$K$&$T$&$K$&$T$&$K$&$T$&$K$&$T$\\
		\hline
		35&$6$&$1718$&$9$&$<1.0$&$4$&$1042$&$6$&$<1.0$&$4$&$425$&$5$&$<1.0$\\
		40&$7$&$2709$&$11$&$<1.0$&$6$&$1434$&$9$&$<1.0$&$4$&$531$&$8$&$<1.0$\\
		45&$9$&$3595$&$14$&$<1.0$&$7$&$1745$&$11$&$<1.0$&$5$&$789$&$10$&$<1.0$\\
		50&$9$&$6482$&$15$&$<1.0$&$8$&$2487$&$12$&$<1.0$&$6$&$1165$&$12$&$<1.0$\\
		55&$10$&$7848$&$16$&$<1.0$&$8$&$4907$&$13$&$<1.0$&$8$&$2698$&$19$&$<1.0$\\
		& & & & & & & & & & & & \\
		100&$-$&$-$&$28$&$31$&$-$&$-$&$22$&$31$&$-$&$-$&$19$&$31$\\
		150&$-$&$-$&$42$&$125$&$-$&$-$&$30$&$88$&$-$&$-$&$22$&$71$\\
		200&$-$&$-$&$51$&$259$&$-$&$-$&$37$&$203$&$-$&$-$&$28$&$165$\\
		250&$-$&$-$&$65$&$542$&$-$&$-$&$48$&$415$&$-$&$-$&$37$&$324$\\
		300&$-$&$-$&$79$&$998$&$-$&$-$&$61$&$767$&$-$&$-$&$45$&$327$\\
		\hline
	\end{tabular}
\end{table*}
\begin{table*}
	\small
	\caption{The solution $K$ and the algorithm runtime $T$ (ms), OPT with MKC}	\label{table:T3}
	\centering
	\begin{tabular}{l|ll|ll|ll|ll|ll|ll}
		\hline
		\multirow{3}{*}{n}
		&\multicolumn{4}{c|}{D=20}&\multicolumn{4}{c|}{D=25}&\multicolumn{4}{c}{D=30}\\
		\cline{2-5}\cline{6-9}\cline{10-13}
		&\multicolumn{2}{l|}{OPT}&\multicolumn{2}{l|}{MKC}&\multicolumn{2}{l|}{OPT} &\multicolumn{2}{l|}{MKC}&\multicolumn{2}{l|}{OPT}&\multicolumn{2}{l}{MKC}\\
		\cline{2-3}\cline{4-5}\cline{6-7}\cline{8-9}\cline{10-11}\cline{12-13}
		&$K$&$T$&$K$&$T$&$K$&$T$&$K$&$T$&$K$&$T$&$K$&$T$\\
		\hline
		35&$9$&$1295$&$12$&$<1.0$&$5$&$764$&$7$&$<1.0$&$4$&$425$&$5$&$<1.0$\\
		40&$10$&$1589$&$13$&$<1.0$&$5$&$948$&$7$&$<1.0$&$4$&$531$&$5$&$<1.0$\\
		45&$10$&$2793$&$14$&$<1.0$&$6$&$1231$&$8$&$<1.0$&$5$&$789$&$6$&$<1.0$\\
		50&$11$&$3687$&$15$&$<1.0$&$7$&$2074$&$9$&$<1.0$&$6$&$1165$&$7$&$<1.0$\\
		55&$12$&$4950$&$16$&$<1.0$&$7$&$3159$&$9$&$<1.0$&$8$&$2698$&$7$&$<1.0$\\
		& & & & & & & & & & & & \\
		100&$-$&$-$&$37$&$16$&$-$&$-$&$23$&$16$&$-$&$-$&$17$&$15$\\
		150&$-$&$-$&$45$&$46$&$-$&$-$&$30$&$31$&$-$&$-$&$18$&$31$\\
		200&$-$&$-$&$56$&$109$&$-$&$-$&$39$&$109$&$-$&$-$&$26$&$94$\\
		250&$-$&$-$&$75$&$218$&$-$&$-$&$50$&$187$&$-$&$-$&$32$&$171$\\
		300&$-$&$-$&$97$&$375$&$-$&$-$&$65$&$343$&$-$&$-$&$45$&$327$\\
		\hline
	\end{tabular}
\end{table*}

In the first experiment, we observe the number of cloudlet $K$ with changing average delay requirments, \textit{i.e.} a sequence of values $D$. Fig. \ref{fig:KvsAvgDudc} depicts the curves of the cloudlet number $K$ obtained by the execution of algorithms MKC, Top-K, Random and the OPT. For all algorithms, with the growing of average cloudlet access delay requirment $D$,  $K$ will decrease, which means algorithm MKC outperforms algorithms Random and Top-K by about $27.6\%$ and $20.1\%$ on average and MKC achieves a near optimal cloudlet number that are no more than $1.2$ times greater as much as the optimal one. Fig. \ref{fig:KvsAvgDwdcc} obtains similar results as Fig. \ref{fig:KvsAvgDudc}. MKH outperforms algorithms Random and Top-K by about $39.6\%$ and $50.1\%$ and achieves a near optimal cloudlet number that are no more than $1.6$ times. Table \ref{table:T2} and Table \ref{table:T3} show the quality and the runtime of the proposed algorithms on the DBOCP problem with different $n$. In the tables, the OPT can generate exact solution, but it cannot work due to the computer memory limit, shown as `−-' in the two tables, with the increase of AP numbers. Therefore, it is reasonable to deduce that, when network size is large, MKC and MKH can continue and still produce good approximate solutions.

\begin{figure}[t]
    \centering
            \includegraphics[width=3.5in]{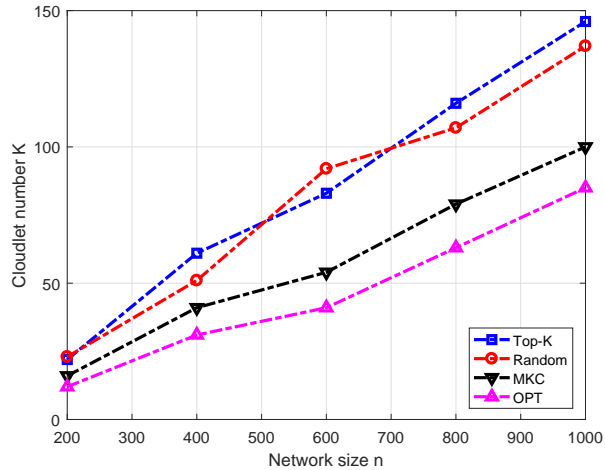}
       \caption{Impact of network size $n$ on the running time of different algorithms w/o designated capacity.}    \label{fig:kvsnwocceps}
\end{figure}

\begin{figure}[t]
	\centering
	\includegraphics[width=3.5in]{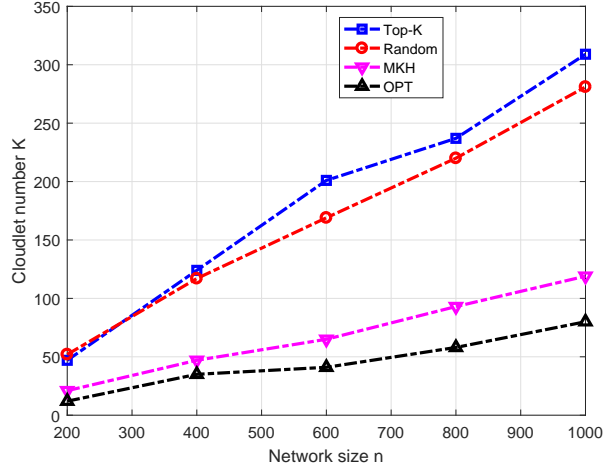}
	\caption{Impact of network size $n$ on the running time of different algorithms w/ designated capacity.}    \label{fig:kvsnwdc}
\end{figure}

Then we evaluate the performance of proposed algorithms MKC and MKH in large networks, by varying $n$ from $200$ to $1000$, while fixing $D$ as $D=30$. Fig. \ref{fig:kvsnwocceps} shows with $n$ increases, $K$ will increase almost linearly. It can be seen that algorithm MKC outperforms algorithms Random and Top-K and Top-K is marginally better than Random. Specifically, MKC is close to the optimal algorithm OPT and it outperforms Top-K as well as Random by about $40\%$ on the deploy cost, \textit{i.e.} the number of deployed cloudlet $K$. Fig. \ref{fig:kvsnwdc} shows the results when cloudlets have identical capacity $c_i=\gamma_{sum}/K$. It can be seen that the average number of used cloudlet by algorithm MKH is less than those of algorithms Random and Top-K by about $57.1\%$ and $60.3\%$, respectively.

\begin{figure}[t]
    \centering
    \includegraphics[width=3.5in]{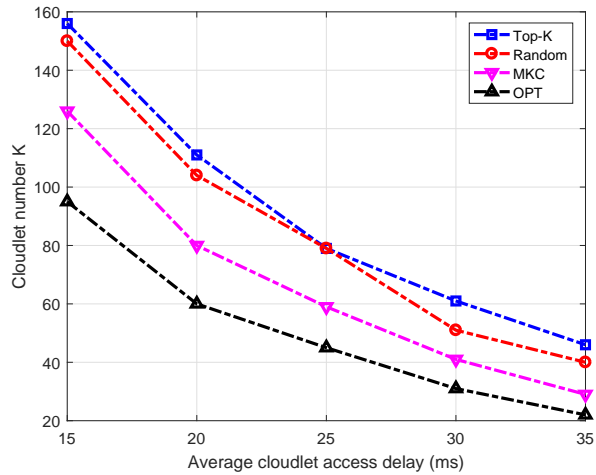}
 \caption{$K$ with required average access delay for un-desiganted capacity.}  \label{fig:kvsdwconsnnocc}
\end{figure}

\begin{figure}[t]
	\centering
	\includegraphics[width=3.5in]{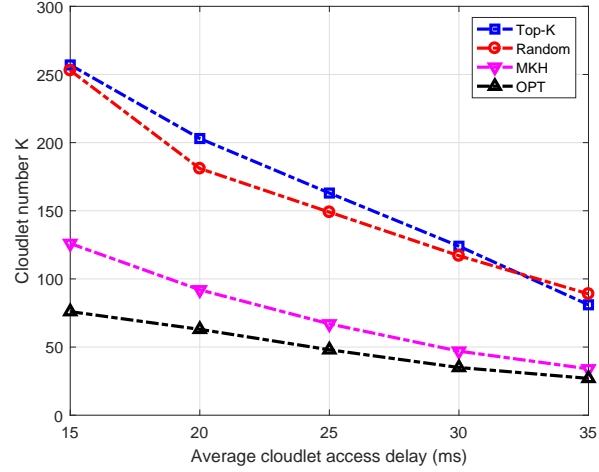}\caption{$K$ with average access delay with identical cloudlet capacities.}
     \label{fig:kvsavgDwcc}
\end{figure}
 
In the final experiment, we study the impact of $K$ with $D$ in the designated capacity case, by varying $D$ from 15 to 35 and fixing $n=400$. Figures \ref{fig:kvsdwconsnnocc} and \ref{fig:kvsavgDwcc} plot the curves of the impact of the average cloudlet access delays. In Fig. \ref{fig:kvsdwconsnnocc}, with the growing number of $D$, $K$ decreases fast. MKC outperforms Random and Top-K by at least $16.7\%$. Figure \ref{fig:kvsavgDwcc} shows the curves of $K$ delivered by different algorithms with identical cloudlet capacity and the required  number of cloudlets by MKH is over $50\%$ less than those by Random and Top-K. MKH algorithm thus is better than Random and Top-K.

\section{Conclusion}\label{sec:conclude}
We have studied the QOECP and DBOCP problems in a WMAN. For the QOECP problem, we have presented MDC and MDE algorithms to improve the performance of existing heuristic algorithm on average cloudlet access delay. For the DBOCP problem, when the network size is small, we devise a K-Mediods based algorithm MKC. For a special case of all cloudlets' computing capabilities have been given, we propose an efficient heuristic algorithm MKH for it. Finally, we evaluated the performance of the proposed algorithms by extensive experimental simulations. Simulation results show that the proposed algorithms are more efficient than existing algorithms on the average cloudlet access delay. Comparing with the random algorithm, the proposed algorithms can reduce the number of deployed cloudlets by at least $16.7\%$. In the future, we will investigate incentive mechanisms to solve the above problems. We will also devise online algorithms to deal with dynamic user traffic.

\section{acknowledgement}
This work was partially supported by the National Key R\&D Probram
of China under Grant No. 2018YFB1003201. It was also supported by
the National Natural Science Foundation of China (NSFC) under Grant
Nos. 61702115 and 61672171. Part of the work was funded by China
Postdoctoral Science Foundation under Grant No. 2017M622632.



\begin{thebibliography}{10}
	\bibitem{pang2015survey}
	Z.~Pang, L.~Sun, Z.~Wang, E.~Tian, and S.~Yang, ``A survey of cloudlet based
	mobile computing,'' in \emph{International Conference on Cloud Computing and
		Big Data}.\hskip 1em plus 0.5em minus 0.4em\relax Shanghai, China: IEEE, 4-6
	Nov. 2015, pp. 268--275.
	\bibitem{zhang2012offload}
	Y.~Zhang, H.~Liu, L.~Jiao, and X.~Fu, ``To offload or not to offload: an
	efficient code partition algorithm for mobile cloud computing,'' in \emph{the
		1st International Conference on Cloud Networking (CLOUDNET)}.\hskip 1em plus
	0.5em minus 0.4em\relax Paris, France: IEEE, 28-30 Nov. 2012, pp. 80--86.
	\bibitem{verbelen2012cloudlets}
	T.~Verbelen, P.~Simoens, F.~De~Turck, and B.~Dhoedt, ``Cloudlets: Bringing the
	cloud to the mobile user,'' in \emph{Proceedings of the third ACM workshop on
		Mobile cloud computing and services}.\hskip 1em plus 0.5em minus 0.4em\relax
	Low Wood Bay, UK: ACM, 25 June 2012, pp. 29--36.
	\bibitem{satyanarayanan2009case}
	M.~Satyanarayanan, P.~Bahl, R.~Caceres, and N.~Davies, ``The case for vm-based
	cloudlets in mobile computing,'' \emph{IEEE pervasive Computing}, vol.~8,
	no.~4, 2009.
	\bibitem{fazio2017prediction}
	P.~Fazio, F.~De~Rango, and M.~Tropea, ``Prediction and qos enhancement in new
	generation cellular networks with mobile hosts: A survey on different
	protocols and conventional/unconventional approaches,'' \emph{IEEE
		Communications Surveys \&amp; Tutorials}, 2017.
	\bibitem{xu2016efficient}
	Z.~Xu, W.~Liang, W.~Xu, M.~Jia, and S.~Guo, ``Efficient algorithms for
	capacitated cloudlet placements,'' \emph{IEEE Transactions on Parallel and
		Distributed Systems}, vol.~27, no.~10, pp. 2866--2880, 2016.
	\bibitem{barioni2008accelerating}
	M.~C.~N. Barioni, H.~L. Razente, A.~J. Traina, and C.~Traina, ``Accelerating
	k-medoid-based algorithms through metric access methods,'' \emph{Journal of
		Systems and Software}, vol.~81, no.~3, pp. 343--355, 2008.
	\bibitem{Ren2014Dynamic}
	S.~Ren and M.~V.~D. Schaar, ``Dynamic scheduling and pricing in wireless cloud
	computing,'' \emph{IEEE Transactions on Mobile Computing}, vol.~13, no.~10,
	pp. 2283--2292, 2014.
	\bibitem{Gu2014Optimal}
	L.~Gu, D.~Zeng, A.~Barnawi, S.~Guo, and I.~Stojmenovic, ``Optimal task
	placement with qos constraints in geo-distributed data centers using dvfs,''
	\emph{IEEE Transactions on Computers}, vol.~64, no.~7, pp. 2049--2059, 2014.
	\bibitem{kosta2012thinkair}
	S.~Kosta, A.~Aucinas, P.~Hui, R.~Mortier, and X.~Zhang, ``Thinkair: Dynamic
	resource allocation and parallel execution in the cloud for mobile code
	offloading,'' in \emph{Proceedings of Infocom}.\hskip 1em plus 0.5em minus
	0.4em\relax IEEE, 2012, pp. 945--953.
	\bibitem{Xu2015Capacitated}
	Z.~Xu, W.~Liang, W.~Xu, and M.~Jia, ``Capacitated cloudlet placements in
	wireless metropolitan area networks,'' in \emph{Local Computer Networks},
	2015, pp. 570--578.
	\bibitem{jia2015optimal}
	M.~Jia, J.~Cao, and W.~Liang, ``Optimal cloudlet placement and user to cloudlet
	allocation in wireless metropolitan area networks,'' \emph{IEEE Transactions
		on Cloud Computing}, 2015.
	\bibitem{cai2014cloudlet}
	W.~Cai, V.~C. Leung, and L.~Hu, ``A cloudlet-assisted multiplayer cloud gaming
	system,'' \emph{Mobile Networks and Applications}, vol.~19, no.~2, pp.
	144--152, 2014.
	\bibitem{bourdena2015using}
	A.~Bourdena, C.~Mavromoustakis, G.~Mastorakis, J.~Rodrigues, and C.~Dobre,
	``Using socio-spatial context in mobile cloud offload process for energy
	conservation in wireless devices,'' \emph{IEEE Transactions on Cloud
		Computing}, 2015.
	\bibitem{hoang2012optimal}
	D.~T. Hoang, D.~Niyato, and P.~Wang, ``Optimal admission control policy for
	mobile cloud computing hotspot with cloudlet,'' in \emph{Wireless
		Communications and Networking Conference (WCNC)}.\hskip 1em plus 0.5em minus
	0.4em\relax Shanghai, China: IEEE, 2012, pp. 3145--3149.
	\bibitem{Cormen2001Introduction}
	T.~H. Cormen, C.~E. Leiserson, R.~L. Rivest, and C.~Stein, ``Introduction to
	algorithms second edition,'' 2001.
	\bibitem{Charikar1999A}
	M.~Charikar, S.~Guha, Tardos, and D.~B. Shmoys, ``A constant-factor
	approximation algorithm for the k -median problem (extended abstract),'' in
	\emph{ACM Symposium on Theory of Computing}, 1999, pp. 1--10.
	\bibitem{Fan2017Cost}
	Q.~Fan and N.~Ansari, ``Cost aware cloudlet placement for big data processing
	at the edge,'' in \emph{IEEE International Conference on Communications},
	2017, pp. 1--6.
	\bibitem{Jin2016Auction}
	A.~Jin, W.~Song, P.~Wang, D.~Niyato, and P.~Ju, ``Auction mechanisms toward
	efficient resource sharing for cloudlets in mobile cloud computing,''
	\emph{IEEE Transactions on Services Computing}, vol.~9, no.~6, pp. 895--909,
	2016.
	\bibitem{gtitm}
	``Gt-itm,'' \url{http://www.cc.gatech.edu/projects/gtitm/}, 2017, [Online;
	accessed 10-May-2017].
	\bibitem{IEEEstd8021990}
	``Ieee standards for local and metropolitan area networks: Overview and
	architecture (ansi),'' 1990.
	\bibitem{lpsolve}
	``Lp-solve,'' \url{http://lpsolve.sourceforge.net}, 2003.
\end{thebibliography}
\end{document}